\documentclass[11pt,a4paper]{article}
\usepackage{epsfig}
\usepackage{amssymb}
\usepackage{here}
\usepackage{cite}
\usepackage{rotating}
\begin{document}
%
%
\newcommand{\PPEnum}    {CERN-EP-2002-XXX}
\newcommand{\PRnum}     {OPAL PR-xxx}
\newcommand{\PNnum}     {OPAL Physics Note PN-xxx}
\newcommand{\TNnum}     {OPAL Technical Note TN-xxx}
\newcommand{\Date}      {12 Dec 2002}
\newcommand{\Author}    {P.~Amaral, M.~Oreglia}
\newcommand{\MailAddr}  {Pedro.Amaral@cern.ch, Mark.Oreglia@cern.ch}
\newcommand{\PublicReading}  {Public Reading: Monday, 9 Dec, 16h00}
\newcommand{\EdBoard}   {D Charlton, J McKenna, T Schoerner, W Scott}
\newcommand{\DraftVer}  {DRAFT 3.2 - FINAL DISPATCH VERSION}
\newcommand{\DraftDate} {\Date}
\newcommand{\TimeLimit} {{\bf Wednesday, 18 December, 9h00 Geneva time}}

\def\toprule{\noalign{\hrule \medskip}}
\def\midrule{\noalign{\medskip\hrule }}
\def\botrule{\noalign{\medskip\hrule }}
\setlength{\parskip}{\medskipamount}

%
\newcommand{\general}  {{\it general}}
\newcommand{\SM}  {{Standard Model}}

\newcommand{\ellell}   {\ell^+ \ell^-}
\newcommand{\ee}       {\mbox{${\mathrm{e}}^+ {\mathrm{e}}^-$}}
\newcommand{\epem}     {{\mathrm e}^+ {\mathrm e}^-}
\newcommand {\mm}      {\mu^+ \mu^-}
\newcommand{\mupair}   {\mbox{$\mu^+\mu^-$}}
\newcommand{\nunu}     {\nu \bar{\nu}}
\newcommand{\tautau}   {\mbox{$\tau^+\tau^-$}}
\newcommand{\qqbar}    {{\mathrm q}\bar{\mathrm q}}
\newcommand{\qb}{\mathrm{b}}
\newcommand{\qpair}    {\mbox{${\mathrm q}\overline{\mathrm q}$}}
\newcommand{\ff}       {{\mathrm f} \bar{\mathrm f}}
\newcommand{\gaga}     {\gamma\gamma}
\newcommand{\WW}       {{\mathrm W}^+{\mathrm W}^-}
\newcommand{\ZZ}       {{\mathrm Z}^{0}{\mathrm Z}^{0}}
\newcommand{\Mrec}      {M_{\mrm{recoil}}}
\newcommand{\mgg}       {$m_{\gamma \gamma}$}
\newcommand{\mdip}      {m_{\gamma \gamma}}
\newcommand{\MZ}        {M_{\mathrm Z}}
\newcommand{\MH}        {M_{\mathrm H}}
\newcommand{\MX}        {M_{\mathrm{X}}}
\newcommand{\MY}        {M_{\mathrm{Y}}}
\newcommand {\ho}        {\mbox{$\mathrm{h}^{0}$}}
\newcommand {\Ao}        {\mbox{$\mathrm{A}^{0}$}}
\newcommand {\Ho}        {\mbox{$\mathrm{H}^{0}$}}
\newcommand {\Zo}        {\mbox{$\mathrm{Z}^{0}$}}
\newcommand {\Zboson}        {{\mathrm Z}^{0}}
\newcommand {\Wpm}           {{\mathrm W}^{\pm}}
\newcommand {\hboson}        {{\mathrm h}^{0}}
\newcommand{\ggjj}{\mbox{$\gaga\;{\rm jet-jet}$ }}
\newcommand{\ggqq}{\mbox{$\gaga {\rm q\bar{q}}$ }}
\newcommand{\ggll}{\mbox{$\gaga\ell^{+}\ell^{-}$}}
\newcommand{\ggnn}{\mbox{$\gaga\nu\bar{\nu}$}}
\newcommand{\degree}    {^\circ}
\newcommand{\roots}       {\sqrt{s}}
\newcommand{\Ecm}         {\mbox{$E_{\mathrm{cm}}$}}
\newcommand{\Egam}        {\mbox{$E_{\gamma}$}}
\newcommand{\EgamA}        {\mbox{$E_{\gamma1}$}}
\newcommand{\EgamB}        {\mbox{$E_{\gamma2}$}}
\newcommand{\Ebeam}       {E_{\mathrm{beam}}} 
\newcommand{\ipb}         {\mbox{pb$^{-1}$}}
\newcommand{\Evis}      {\mbox{$E_{\mathrm{vis}}$}}
\newcommand{\Rvis}      {\mbox{$R_{\mathrm{vis}}$}}
\newcommand{\Mvis}      {\mbox{$M_{\mathrm{vis}}$}}
\newcommand{\Rbal}      {\mbox{$R_{\mathrm{bal}}$}}
\newcommand{\onecol}[2] {\multicolumn{1}{#1}{#2}}
\newcommand{\colcen}[1] {\multicolumn{1}{|c|}{#1}}
\newcommand{\ra}        {\rightarrow}   
\newcommand{\ov}        {\overline}   
\def\mrm       {\mathrm}
\newcommand{\Z}{${\rm Z}$ }
\newcommand{\Zbb}{${\rm Z} \rightarrow {\rm b} \overline{\rm b}$ }
\newcommand{\evis}{E_{\rm vis}^{\rm hemi}}
\newcommand{\emis}{E_{\rm miss}^{\rm hemi}}
\newcommand{\ebeam}{E_{\rm beam}}
\newcommand{\ecorr}{E_{\rm corr}}
\newcommand{\ccbar}{\ifmmode {\rm c}\bar{\rm c} \else c$\bar{\mbox{\rm c}}$\fi}
\newcommand{\bbbar}{\ifmmode {\rm b}\bar{\rm b} \else b$\bar{\mbox{\rm b}}$\fi}
\newcommand{\dedx}{\ifmmode {\rm d}E/{\rm d}x \else d$E$/d$x$\fi}
\newcommand{\Dzero}{\ifmmode {\rm D}^0 \else D$^0$\fi}
\newcommand{\Dstar}{\ifmmode {\rm D}^* \else D$^*$\fi}
\newcommand{\Dstarp}{\ifmmode {\rm D}^{*+} \else D$^{*+}$\fi}
\newcommand{\Dss}{\ifmmode {\rm D}^{**} \else D$^{**}$\fi}
\newcommand{\Dssz}{\ifmmode {\rm D}^{**0} \else D$^{**0}$\fi}
\newcommand{\Dssc}{\ifmmode {\rm D}^{**+} \else D$^{**+}$\fi}
\newcommand{\dst}{\ifmmode {\rm D}_{\rm s}^{-} \else D$_{\rm s}^{-}$\fi}
\newcommand{\BR}{\ifmmode {\mathrm{BR}} \else BR\fi}
\newcommand{\myb}{\ifmmode {\mathrm{b}} \else b\fi}
\newcommand{\mBR}{\ifmmode {\mathrm{BR}} \else BR\fi}
\newcommand{\dstaunu}{\mbox{D$_{\rm s}^- \rightarrow \tau^- \overline{\nu}_\tau$}}
\newcommand{\brjd}{\mbox{$\BR\left(\myb\ \rightarrow B^-\right) \times \BR\left(B^- \rightarrow D^{0}_{1}\, \ell^- \bar{\nu} \right)\ \times \BR\left(D^{0}_{1} \rightarrow \Dstarp\pi^{-}\right)$}}
\newcommand{\newbrjd}{\mbox{$\BR\left(\myb\ \rightarrow \bar{\mathrm{B}} \right) \times \BR\left(\bar{\mathrm{B}} \rightarrow \mathrm{D}^{0}_{1}\, \ell^- \bar{\nu} X \right)\ \times \BR\left(\mathrm{D}^{0}_{1} \rightarrow \Dstarp\pi^{-}\right)$}}
\newcommand{\brj}{\mbox{$\BR\left(b \rightarrow B^-\right) \times \BR\left(B^- \rightarrow \Dssz\, \ell^- \bar{\nu} \right)\ \times \BR\left(\Dssz \rightarrow \Dstarp\pi^{**(-)}\right)$}}
\newcommand{\brdssds}{\mbox{$\BR\left(\Dssz \rightarrow \Dstarp\pi^{**(-)}\right)$}}
\newcommand{\brdsd}{\mbox{$\BR\left(\Dstarp \rightarrow \Dzero\pi^{*+}\right)$}}
\newcommand{\brdkpi}{\mbox{$\BR\left(\Dzero \to K^+\pi^-\right)$}}
\newcommand{\brdktpi}{\mbox{$\BR\left(\Dzero \to K\,3\pi\right)$}}
\newcommand{\dssds}{\mbox{$\Dssz \rightarrow \Dstarp\pi^{-}$}}
\newcommand{\dsd}{\mbox{$\Dstarp \rightarrow \Dzero\pi^{+}$}}
\newcommand{\dkpi}{\mbox{$\Dzero \to \mathrm{K}\pi$}}
\newcommand{\dktpi}{\mbox{$\Dzero \to \mathrm{K}\,3\pi$}}
\newcommand{\dktpiorkpi}{\mbox{$\Dzero \to \left(\mathrm{K}\pi\: \mathrm{or}\, \mathrm{K}\,3\pi\right)$}}
\newcommand{\dktpicommakpi}{\mbox{$\Dzero \to \mathrm{K}\pi , \, \mathrm{K}\,3\pi$}}
\newcommand{\kpi}{\mbox{$K\pi$}}
\newcommand{\ktpi}{\mbox{$K\,3\pi$}}
\newcommand{\kpipizero}{\mbox{$K\pi\pi^{0}$}}
\newcommand{\backform}{\mbox{$x^{\gamma}e^{-\beta x}$}}

\newcommand{\bj}{B^- \to \Dssz \ell^- \bar{\nu}}
\newcommand{\bjx}{\bar{\mathrm{B}} \to \Dss \ell^- \bar{\nu} X}
\newcommand{\bjxn}{\bar{\mathrm{B}} \to \Dssz \ell^- \bar{\nu} X}
\newcommand{\bjxc}{\bar{\mathrm{B}} \to \Dssc \ell^- \bar{\nu} X}
\newcommand{\bdss}{\bar{\mathrm{B}}^0 \to \Dstarp \ell^- \bar{\nu}}
\newcommand{\bbar}{\mbox{$\bar{b}$}}
\newcommand{\piss}{\mbox{$\pi^{**}$}}
\newcommand{\pislow}{\mbox{$\pi_{\mathrm{slow}}$}}
\newcommand{\dms} {\mbox{$\Delta m^{*}$}}
\newcommand{\dmss} {\mbox{$\Delta m^{**}$}}
\newcommand{\dssm} {\mbox{$\Delta m^{**}$}}
\newcommand{\chisquare} {\mbox{$\chi^{2}$}}

%
%
\newcommand{\gsim}{\;\raisebox{-0.9ex}
           {$\textstyle\stackrel{\textstyle >}{\sim}$}\;}
%
%
%
\newcommand{\PhysLett}  {Phys.~Lett.}
\newcommand{\PRL}       {Phys.~Rev.\ Lett.}
\newcommand{\PhysRep}   {Phys.~Rep.}
\newcommand{\PhysRev}   {Phys.~Rev.}
\newcommand{\NPhys}     {Nucl.~Phys.}
\def\NIM                {\mbox{Nucl. Instr. Meth.}}
\newcommand{\NIMA}[1]   {\NIM\ {\bf A{#1}}}
\newcommand{\IEEENS}    {IEEE Trans.\ Nucl.~Sci.}
\newcommand{\ZPhysC}[1]    {Z. Phys. {\bf C#1}}
\newcommand{\EurPhysC}[1]    {Eur. Phys. J. {\bf C#1}}
\newcommand{\PhysLettB}[1] {Phys. Lett. {\bf B#1}}
\newcommand{\CPC}[1]      {Comp.\ Phys.\ Comm.\ {\bf #1}}
\def\etal{\mbox{{\it et al.}}}
\newcommand{\JPG}[3] {J.~Phys.\ {G#1} (#2) #3}
%
%
\newcommand{\OPALColl}  {OPAL Collaboration}
\newcommand{\JADEColl}  {JADE Collaboration}
%
\newcommand{\ntotaabout}{$3.90$ }
\newcommand{\ntot}{$3\,904\,417$ }


\begin{titlepage}
\begin{center}{\large   EUROPEAN ORGANIZATION FOR NUCLEAR RESEARCH
}\end{center}\bigskip
\begin{flushright}
       CERN-EP-2002-094  \\ 20 December 2002
\end{flushright}
\bigskip\bigskip\bigskip\bigskip\bigskip
%
%
\begin{center}{\huge\bf\boldmath A Measurement of \\ 
                                 Semileptonic B Decays to Narrow \\ 
                                 Orbitally-Excited Charm Mesons\\
                                 }
\end{center}\bigskip\bigskip

\begin{center}{\LARGE The OPAL Collaboration
}\end{center}\bigskip\bigskip
%
%
\bigskip\begin{center}{\large  Abstract}\end{center}
The decay chain 
b $ \rightarrow \bar{\mathrm{B}} \rightarrow \Dssz\, \ell^-
\bar{\nu} X $, $\Dssz \rightarrow \Dstarp\pi^{-}$, $\dsd$, $\dktpiorkpi$ 
is identified in a sample of 3.9 million hadronic Z decays collected with the OPAL detector at LEP. 
The branching ratio \sloppy{$\newbrjd$} is measured to be $(2.64 \pm
0.79\,(\mathrm{stat}) \pm 0.39\,(\mathrm{syst})) \times 10^{-3}$ 
for the $J^{P}=1^{+}$ (D$^{0}_{1}$) state. 
For decays into the $J^{P}=2^{+}$ (D$_{2}^{*0}$) state, 
an upper limit of $1.4 \times 10^{-3}$ is placed on the branching ratio at the 95\% confidence level.

\bigskip\bigskip\bigskip\bigskip

\begin{center}{\large
To be submitted to European Journal of Physics C \\

}\end{center}

\end{titlepage}
\begin{center}{\Large        The OPAL Collaboration
}\end{center}\bigskip
\begin{center}{
G.\thinspace Abbiendi$^{  2}$,
C.\thinspace Ainsley$^{  5}$,
P.F.\thinspace {\AA}kesson$^{  3}$,
G.\thinspace Alexander$^{ 22}$,
J.\thinspace Allison$^{ 16}$,
P.\thinspace Amaral$^{  9,  aa}$, 
G.\thinspace Anagnostou$^{  1}$,
K.J.\thinspace Anderson$^{  9}$,
S.\thinspace Arcelli$^{  2}$,
S.\thinspace Asai$^{ 23}$
% ...
}\end{center}\bigskip
\bigskip
$^{  1}$School of Physics and Astronomy, University of Birmingham,
Birmingham B15 2TT, UK
\newline
$^{  2}$Dipartimento di Fisica dell' Universit\`a di Bologna and INFN,
I-40126 Bologna, Italy
\newline
$^{  3}$Physikalisches Institut, Universit\"at Bonn,
D-53115 Bonn, Germany
\newline
% ...
$^{  5}$Cavendish Laboratory, Cambridge CB3 0HE, UK
\newline
% ...
$^{  9}$Enrico Fermi Institute and Department of Physics,
University of Chicago, Chicago IL 60637, USA
\newline
% ...
$^{ 16}$Department of Physics, Schuster Laboratory, The University,
Manchester M13 9PL, UK
\newline
% ...
$^{ 22}$Department of Physics and Astronomy, Tel Aviv University,
Tel Aviv 69978, Israel
\newline
$^{ 23}$International Centre for Elementary Particle Physics and
Department of Physics, University of Tokyo, Tokyo 113-0033, and
Kobe University, Kobe 657-8501, Japan
\newline
% ...
$^{  b}$ and Royal Society University Research Fellow
\newline
$^{  c}$ and Institute of Nuclear Research, Debrecen, Hungary
\newline
$^{  d}$ and Heisenberg Fellow
\newline
$^{  e}$ and Department of Experimental Physics, Lajos Kossuth University,
 Debrecen, Hungary
\newline
$^{  f}$ and MPI M\"unchen
\newline
% ...
$^{  h}$ now at University of Liverpool, Dept of Physics,
Liverpool L69 3BX, U.K.
\newline
% ...
$^{  q}$ now at IPHE Universit\'e de Lausanne, CH-1015 Lausanne, Switzerland
\newline
% ...
$^{  s}$ now at Universitaire Instelling Antwerpen, Physics Department, 
B-2610 Antwerpen, Belgium
\newline
% ...

\newpage
\section{Introduction}
\label{sec:intro}

Semileptonic B decays to orbitally-excited P-wave charm mesons (D$^{**}$) are of interest
for several reasons. Firstly, they constitute a significant fraction of B
semileptonic decays, thereby accounting for some of the difference between the inclusive measurements
and the sum of the exclusive B decay
modes to D$^* \ell \bar{\nu}$ and  D$ \ell \bar{\nu}$~\cite{pdg02,richman,aleph}. 
They also contribute 
the major source of systematic error in the
$ \mathrm{|V_{cb}|}$ measurement at LEP, as a background to the direct decay  
$\bdss$.
Finally, the measured decay properties can be compared with theoretical HQET predictions~\cite{neubert,richman}.

The D$^{**}$ mesons (sometimes denoted $\mathrm{D}_J$) are composed of a charm quark and a
light quark in a state of orbital angular momentum $L=1$.
In the infinite heavy-quark (charm) mass limit the D$^{**}$ system is
equivalent to a fixed force center one-body problem. Hence the total
(spin+orbital) angular momentum of the light quark degrees of freedom 
(labelled by $J_{q}=\frac{1}{2}$ or $\frac{3}{2}$) 
and the spin of the heavy quark are taken as separately conserved~\cite{richman,rosner}.
The 
$J_{q}$=$\frac{3}{2}$ 
states combine with the heavy quark spin to form two states with 
$J^{P}=1^{+}$ (D$_{1}$) and
$J^{P}=2^{+}$ (D$_{2}^{*}$). 
In the infinite charm mass limit, they can only undergo D-wave decay, 
and therefore have narrow widths~\cite{rosner}.
For 
$J_{q}$=$\frac{1}{2}$ 
we have two states with  $J^{P}=0^{+}$ and $J^{P}=1^{+}$. 
The 
$J_{q}$=$\frac{1}{2}$ 
states can decay via S-wave and are expected to be
broad, but 
their experimental observation is still not established~\cite{pdg02}.
So in total, for P-wave mesons, four charged and four neutral D$^{**}$ states are predicted.
Table~\ref{tabledss} summarises the properties of the neutral states.

In this paper we present a new measurement of semileptonic B decays into the narrow
neutral D$_1^0$ and D$_2^{*0}$ states. 
We reconstruct events compatible with the decay chain 
\begin{eqnarray}
\label{eqnchain}
\myb\ \rightarrow \bar{\mathrm{B}} \rightarrow \Dssz\, \ell^- \bar{\nu} X,\, 
\Dssz \rightarrow \Dstarp\pi^{-},\, \dsd,\, \dktpiorkpi.
\end{eqnarray} 
Here, and throughout this paper, $\bar{\mathrm{B}}$ refers to the 
$\bar{\mathrm{B}}^0$ and $\mathrm{B}^{-}$ mesons,
and charge conjugate modes and reactions are always implied.
Also, $\ell$ refers to both electrons and muons, and the terms
``kaon'' and ``pion'' denote the charged particles.

These results update a previous OPAL 
analysis~\cite{oldopal}. 
Similar measurements have been performed also by ALEPH~\cite{aleph},
DELPHI~\cite{delphi}, CLEO~\cite{cleo} and ARGUS~\cite{argus}. 
We note the complementarity between the LEP and 
$\Upsilon(4\mathrm{S})$ analyses,
as in the latter, 
theoretical uncertainties in the B semileptonic decay form factors 
contribute significantly to the overall systematic error.
This is less important at LEP, due to the high boost of the B hadron and its decay products.

\section{Detector and Monte Carlo Samples}
\label{s:det}

The OPAL detector is fully described 
elsewhere~\cite{detector}. A brief description of the main components relevant for 
this analysis follows. Tracking of charged particles is performed by a central detector,
consisting of a silicon microvertex detector, a vertex chamber, a jet chamber
and $z$-chambers\,\footnote{A right handed coordinate system is used, with
positive $z$ along the $\mathrm{e}^-$ beam direction and $x$ pointing
towards the center of the LEP ring. The polar and azimuthal angles are
denoted by $\theta$ and $\phi$, and 
the origin is taken to be the center of the detector.}.
The central detector is inside a
solenoid, which provides a uniform axial magnetic field of 0.435\,T.
The silicon microvertex detector consists of two layers of
silicon strip detectors; for most of the data used in this paper,
the inner layer covered a polar angle range of $|\cos\theta |<0.83$ and
the outer layer covered $| \cos \theta |< 0.77$, with
an extended coverage for the data taken after 1996. This detector 
provided only $\phi$-coordinate information for data taken in 1991--1992, and
also $z$-coordinate information thereafter. The vertex chamber is a precision drift chamber
which covers the range $|\cos \theta | < 0.95$.
The jet chamber is a large-volume drift chamber, 4.0~m long and 3.7~m in diameter,
providing both tracking and ionisation energy loss (d$E$/d$x$) information.
The $z$-chambers measure the $z$-coordinate
of tracks as they leave the jet chamber in the range
$|\cos \theta | < 0.72$. Immediately outside the tracking volume is the 
solenoid and a time-of-flight 
counter array followed by an electromagnetic shower presampler and
a lead-glass electromagnetic calorimeter. 
The return yoke of the magnet lies outside the electromagnetic
calorimeter and is instrumented with limited streamer chambers
and thin gap chambers.
It is used as a hadron calorimeter and assists in the reconstruction
of muons. The outermost part of the detector is made up by
layers of muon chambers.

A Monte Carlo simulation of hadronic Z decays of about five times
the size of the recorded data sample for b flavoured events, and about 
the same size as the recorded data for the other flavours, is used in the 
analysis. In addition, signal Monte Carlo samples
were generated with approximately
one thousand times the expected number of signal events in
the data.
The simulated events were generated using JETSET~7.4~\cite{bib:jetset} 
with the b and c quark fragmentation modelled according to the 
parameterisation of Peterson et al.~\cite{bib:peter}.
A global fit to OPAL 
data has been performed to optimise the JETSET parameters~\cite{bib:tuned}.
These events were processed using the full
OPAL detector simulation~\cite{GOPAL} and analysed in the 
same manner as the data.

\section{Event Selection} 

\begin{figure}[!p]
        \vspace{0.8cm}
        \begin{center}
            \resizebox{\linewidth}{!}{\includegraphics{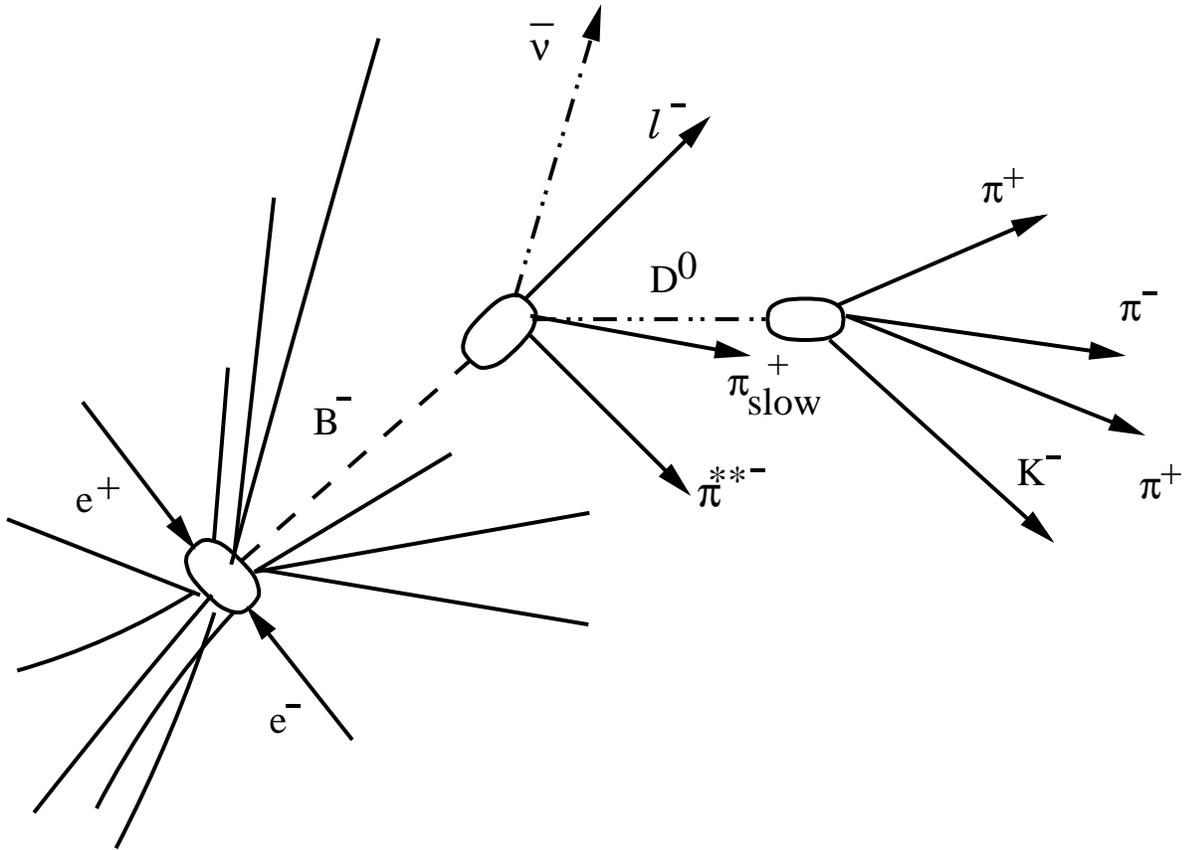} }
        \caption[topology]{    
                 Event topology for a semileptonic B decay into a $\Dssz\, \ell^- \bar{\nu}$, $\Dssz \rightarrow \Dstarp\pi^{-}$, $\dsd$, $\dktpi$,
                 with the 3 reconstructed vertices shown.
        \label{f:topology} }
        \end{center}
    \end{figure}

This analysis identifies decays compatible with the decay chain of Eq.~\ref{eqnchain},
illustrated in Fig.~\ref{f:topology}.
We attempt to identify the 5 relevant tracks in the case of the
$\dkpi$ mode, 
or 7 tracks for the $\dktpi$ mode, that match the correct event topology.
To distinguish between the 
different charged pions involved, we denote as $\piss$ the one coming from the $\Dssz$ decay. 
Similarly, $\pislow$ comes from the D$^{*+}$ meson, 
its name arising from the small mass difference ($\dms$) between the $\Dstarp$ and the $\Dzero$,
leading to a very low pion momentum in the $\Dstarp$ rest frame.
Event selection criteria are applied to the data and Monte Carlo samples in five stages, namely
hadronic Z preselection, reconstruction of $\Dstarp\ell^-$candidates, identification of the
best $\piss$, selection of the best overall candidate, and final background suppression cuts
specific to the $\dkpi$ and $\dktpi$ modes.

\subsection{Hadronic Z Decay Preselection}
\label{s:presel}

Hadronic Z decays collected with the OPAL detector
at e$^+$e$^-$ centre-of-mass energies near the 
Z resonance are selected using a standard OPAL 
hadronic event selection~\cite{TKMH}, 
with an additional requirement of at least 7 tracks per event.
All the tracks must pass a set of standard quality cuts~\cite{bib:gambb}.
The selection efficiency for hadronic Z decays
is $(98.1 \pm 0.5)\%$~\cite{bib:gambb} with a background of $(0.11\pm 0.03)$\%. 
Only data that were taken with the silicon microvertex detector in operation
are used in this analysis.
After this preselection, the resulting data 
sample collected in the years 1991--2000 consists of $\ntot$ events selected from a  
total integrated luminosity of 180.8~$\ipb$ collected at the Z resonance.
About $11\%$ of the data used were recorded in the years 1996--2000.

\subsection{Reconstruction of $\Dstarp\ell^-$ Candidates}
\label{s:reco}

Candidate events must have two or more jets, defined
using a cone algorithm~\cite{ref:cone}, with radius of 0.7 radians and 10 GeV minimum energy.
The primary vertex of the event is reconstructed using a beam spot constraint~\cite{ref:primary}. 
We then identify muon or electron candidates with momentum greater than 3 or 2 GeV, respectively. 
The muon cut is more stringent to reduce the background from 
pions misidentified as muons.
An artificial neural network is used to identify electrons~\cite{ref:NN},
and photon conversions are rejected as described in~\cite{ref:ll}.
Muons are selected as described in~\cite{bib:gambb}.

Next, we look for tracks consistent with the $\Dstarp$ hypothesis and  
within the same jet as the lepton.
Firstly, a kaon candidate with the same charge sign as the lepton,
and momentum greater than 1.0 GeV,
 is required to have a 
\dedx\ probability greater than 1\% 
for the kaon hypothesis. 
This requirement is increased to 10\%
if the $\Dstarp$ energy is smaller than half the beam energy, where
$\mathrm{K/\pi}$ separation
is more powerful.
Also, the measured \dedx\ of the kaon candidate is required to be smaller
than the expected \dedx\ for the pion hypothesis. 

A pion of opposite charge is then sought in order to form a 
$\dkpi$ candidate;
for $\dktpi$, two additional (pion) tracks of opposite charge are identified.
The reconstructed $\Dzero$ mass peak has a resolution of about 25 MeV;
a loose cut of $1.79<M(\Dzero)<1.94$~GeV is applied. A $\pislow$ 
track candidate of opposite charge to the kaon is combined with the $\Dzero$ to form a $\Dstarp$. 
The mass difference $\dms~\equiv~M(\Dstarp)-M(\Dzero)$ forms
a sharp peak of about 1~MeV resolution, and we require 
$140.5<\dms<149.5$~MeV at this stage.
The energy of the $\Dstarp$ is required to be greater than 15\% of the beam energy.
A $\Dzero$ vertex is required to be successfully reconstructed using the 2 (or 4) candidate tracks.
A kinematic fit performed on the $\Dstarp$ tracks
using the known $\Dstarp$ and $\Dzero$ masses as constraints
is required to converge. 
We note that at this point, several different $\Dstarp\ell^-$ candidates per event can exist.

\subsection{Selection of Best Overall Candidate }
\label{s:selectpiss}

We now proceed to select the best $\piss$ candidate for a given $\Dstarp\ell^-$ assignment,
where $\piss$ candidates are required at this stage
to have an energy greater than 0.6 GeV and the same charge as the lepton.
A B vertex is constructed using the tracks of the $\piss$, $\pislow$, lepton, and reconstructed
$\Dzero$. The primary
vertex is then recomputed, excluding all of the above 5 (or 7) tracks. 
The new $\chisquare$ of the 
primary vertex is obtained, and for each $\Dstarp\ell^-$ combination the
$\piss$ candidate that gives the smallest combined
$\mathrm{ \chisquare }$ of the primary and B vertices is chosen.
The B and D decay lengths 
are recomputed with the new vertex positions.
In order to select only well reconstructed B vertices, we require an 
estimated uncertainty 
of less than 1~mm on the B decay length.

Having selected the best $\piss$ candidate for the given $\Dstarp\ell^-$ assignment,
we reconstruct the mass difference 
$\dssm~\equiv~M(\Dssz)-M(\Dstarp)$, and restrict it to the range 0.14--1.10 GeV.
We then select for further analysis only one $\Dstarp\ell^-$ candidate per event, namely
the one with highest $\Dstarp$ kinematic fit probability. 
This cut greatly reduces
the combinatorial background arising from fake combinations
when forming the $\Dstarp\ell^-$.

\subsection{Final Cuts}
\label{s:finalcuts}

The main backgrounds at this point arise from fragmentation, 
where a pion from the b quark fragmentation fakes
a $\piss$ from a $\Dssz$, and combinatorial backgrounds;
we include in the fragmentation category
pions from $\mathrm{B}^{**}\rightarrow\mathrm{B}\pi X$.
The cuts for reducing these backgrounds differ somewhat for 
the two $\Dzero$ decay modes, the $\dktpi$ mode having larger backgrounds, and 
correspondingly stronger cuts.

To reduce the fragmentation background, we make the following requirements:
\begin{itemize}
\setlength{\itemsep}{-3pt}

\item The $\piss$ energy must be greater than 1 GeV.

\item The output of a neural network applied to the $\piss$ track 
is required to be greater than
0.6 for the $\dktpi$ mode, and greater than 
0.3 for the $\dkpi$ mode. 
This neural network uses momentum, transverse
momentum to the jet, and impact parameter significance with respect to the primary vertex;
it is used to distinguish between tracks from the primary vertex and
genuine B decay 
tracks~\cite{ref:btnnps}.

\item The B decay length significance 
$l_\mathrm{B}/\sigma_{l_{\mathrm{B}}}$ is required to be greater than $1.5$, 
where $l_\mathrm{B}$ is the signed B decay length and
$\sigma_{l_{\mathrm{B}}}$ its error. 
It is signed positive if the B vertex is displaced from 
the primary vertex in the same direction as the jet momentum, and negative otherwise. 

\item Similarly, $l_{\mathrm{B}-\mathrm{D}}/\sigma_{l_{\mathrm{B}-\mathrm{D}}}~>-2.0$ 
for the signed decay length significance
 between the B and D vertices. It is signed positive if the D vertex is displaced from 
the B vertex in the same direction as the jet momentum, and negative otherwise.

\end{itemize}

The combinatorial background is suppressed using the following cuts:
\begin{itemize}
\setlength{\itemsep}{-3pt}

\item $143 < \dms < 148$ MeV for $\dktpi$, 
$142 < \dms < 149$ MeV for $\dkpi$.
\item $1.815 < M(\Dzero) < 1.915$ GeV for $\dktpi$.

\item $\cos{\theta^*}$ $>-0.9$, where $\theta^*$ is the angle between the kaon and the 
$\Dzero$ boost direction, calculated in the $\Dzero$ rest frame. This uses the fact
that the $\Dzero$ is a pseudoscalar, whereas the background, particularly that resulting
from particle misidentification, tends to peak at large negative values of $\cos{\theta^*}$.

\item For the $\dktpi$ mode, the product of the 
neural network applied 
to the 4 tracks from the $\Dzero$ decay, is required to be greater than 0.2.
\end{itemize}

Fake leptons and non-$\bar{\mathrm{B}}$ semileptonic decays appear at a smaller rate
than the previous backgrounds,
but their invariant mass distributions can peak in the expected
signal region. 
The non-$\bar{\mathrm{B}}$ semileptonic background arises primarily from charm, tau
and b-baryon decays, and pion misidentification.
Therefore we impose the following additional cuts:
\begin{itemize}
\setlength{\itemsep}{-3pt}

\item The $\Dstarp\ell^-$ invariant mass is restricted to the range 2.8--5.0 GeV.

\item The lepton momentum transverse to the jet:
$\mathrm{p_{T,\ell}}>0.9$~GeV for $\dktpi$, and $>0.6$~GeV for $\dkpi$.
This cut is effective in reducing misidentified lepton backgrounds
and does not reduce the signal significance.

\end{itemize}

\begin{figure}[!p]
        \vspace{0.8cm}
        \begin{center}
            \resizebox{\linewidth}{!}{\includegraphics{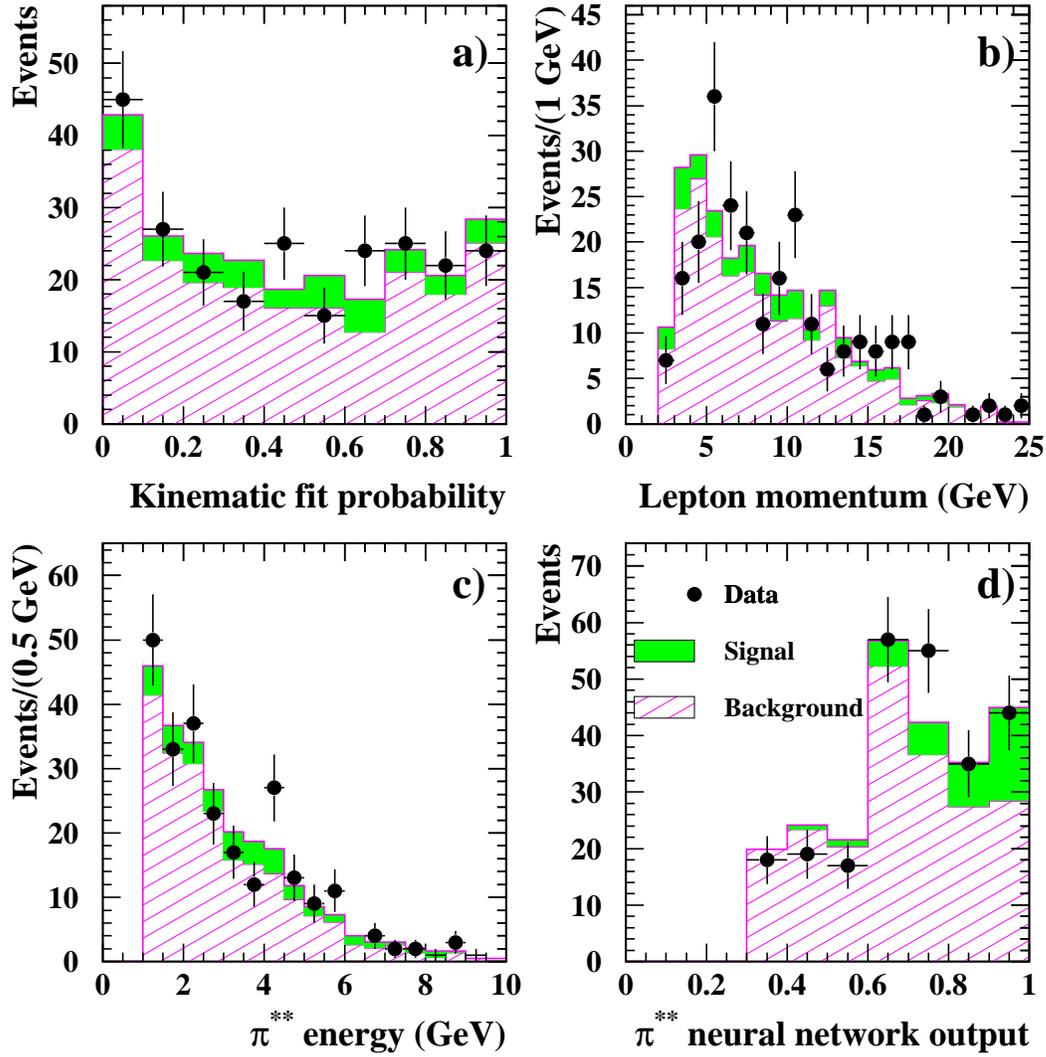} }
        \caption[datamc]{
Distributions of data, simulated signal,
 and backgrounds after all cuts for: 
a) probability of kinematic fit, 
b) lepton momentum, 
c) $\piss$ energy, 
d) $\piss$ neural network output. 
Both $\Dzero$ decay modes were combined, and the signal rate is the one measured in this paper.
        \label{f:datamc} }
        \end{center}
    \end{figure}

These selection criteria were designed to maximise the \emph{efficiency$\times$purity}
using only quantities well modelled by the
Monte Carlo simulation. As an example, Fig.~\ref{f:datamc} presents the data versus Monte Carlo
distributions after all cuts for four of the most relevant quantities used in the analysis. 
Since the final analysis consists of searches for peaks in the distribution of
$\dssm$, the cuts were also tuned
so that the simulated background $\dssm$ distribution
shows no peak in the signal region, $0.35<\dssm<0.55$ GeV.

\section{Backgrounds and Wrong Sign Sample}
\label{s:wsign}

\begin{figure}[!p]
        \vspace{0.8cm}
        \begin{center}
            \resizebox{\linewidth}{!}{\includegraphics{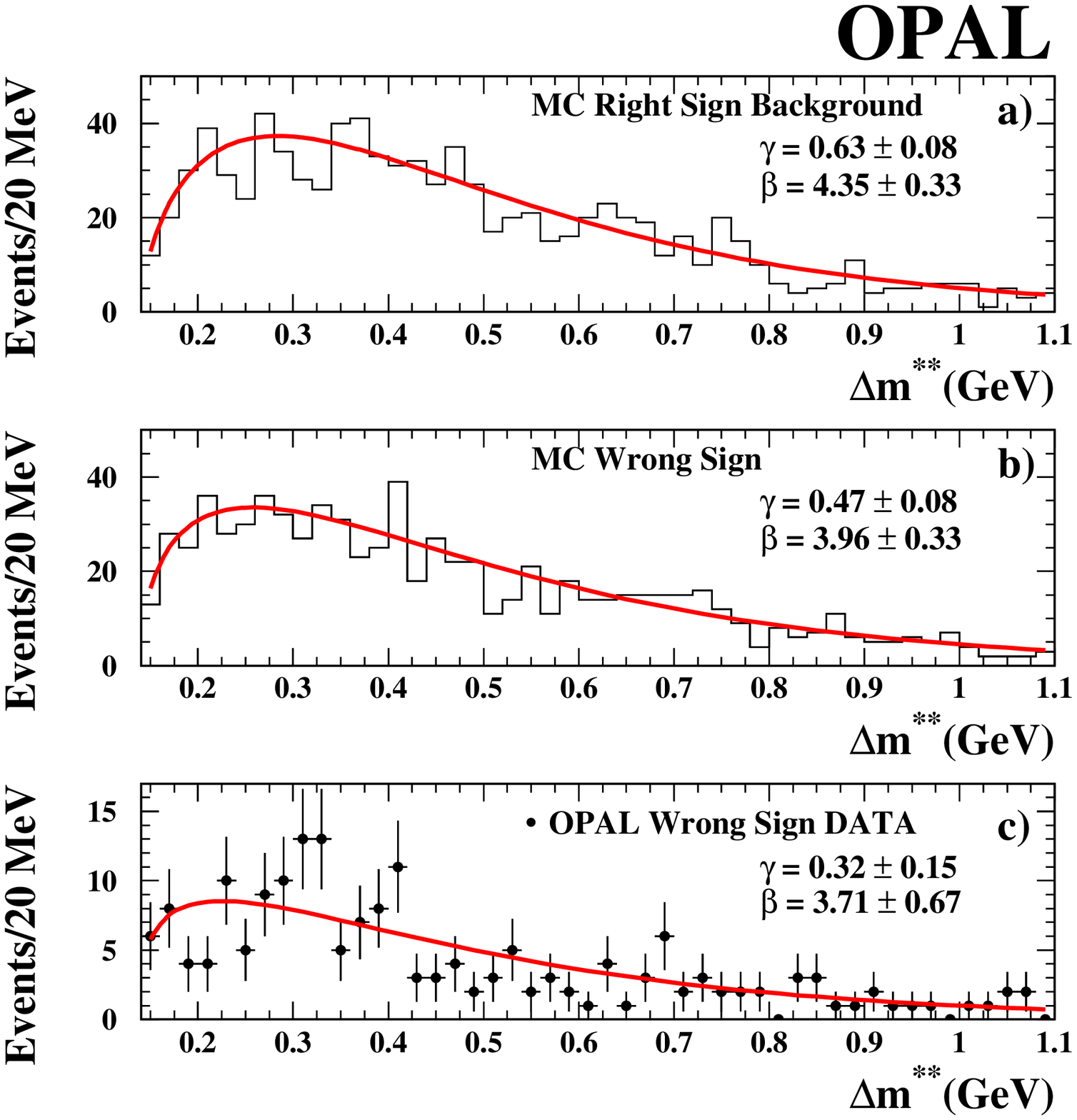} }
        \caption[resolution]{    
$\dssm$ distributions for the combined modes $\dktpiorkpi$: 
a) Monte Carlo background right sign events,
b) Monte Carlo background wrong sign events, 
c) OPAL wrong sign $\dssm$ data.
The superimposed lines are the fits to the functional form $\backform$ with $x=\dssm-m_{\pi}$.
        \label{f:wsign} }
        \end{center}
    \end{figure}

Applying the previous selection to our combined $\dktpiorkpi$ Monte Carlo samples,
the major background process is $\bdss$, 
where the \Dstarp\ is combined with an additional pion of the right sign. 
This comprises 65\% of the total background. The decays $\bjxn$ and $\bjxc$, where
some of the tracks are incorrectly matched, amount to 16\%
(using the branching fractions measured in this paper). 
These three distinct physical
processes have in common that the $\piss$ track is incorrectly identified,
and therefore they constitute the major part of the above mentioned fragmentation background.
The kinematics of this dominant background should be equivalent to the one
obtained by requiring a $\piss$ of the opposite charge from that expected for $\Dssz$ decay.
The Monte Carlo $\dmss$ distributions for background right sign and wrong sign $\piss$ are shown
in Fig.~\ref{f:wsign}a,b. 
They are fitted to a functional form $\backform$ where $x\equiv\dssm-m_{\pi}$; 
the fitted
values of the $\gamma$ and $\beta$ parameters 
for the two samples are consistent. 
Fig.~\ref{f:wsign}c shows the OPAL data wrong sign $\dmss$ distribution;
the fitted $\gamma$ and $\beta$ parameters agree with those obtained in Monte Carlo.

Fake leptons, B$_\mathrm{s}$, and non-$\bar{\mathrm{B}}$ semileptonic decays constitute less
severe backgrounds amounting
to 3\%, 5\% and 11\% of the total background, respectively. We find the contribution
from Z decays into charm and lighter quarks to be negligible.

\section{Signal Fitting Procedure}
\label{s:fit}

An unbinned maximum likelihood fit is performed simultaneously 
on the $\dmss$ right sign and wrong sign distributions.
For the right sign fit,  
the function is given by the sum of two Breit-Wigner distributions, 
each convolved with a Gaussian resolution function, 
plus the background function $\backform$
described previously;
for the wrong sign fit, only the background function is used. 
There are six fitted parameters: 
the number of D$_{1}^{0}$ and D$_2^{*0}$,
the normalisation of the right sign and wrong sign backgrounds, 
and the background shape parameters $\gamma$ and $\beta$.
Thus, the background shape and normalisation are obtained directly from data.
We fix the narrow states masses and widths to the world averages~\cite{pdg02}, 
shown in Table~\ref{tabledss}.
The detector resolution on the reconstructed $\dmss$
is fixed at $\sigma$=8~MeV, as obtained from Monte Carlo.

To check for biases and to assess the systematic uncertainties,
the fitting procedure was tested 
by comparing the results of five simulated experiments, 
each one with the same statistics as the data
(so-called ``ensemble tests''). 
The pull distributions for the fitted number of D$_{1}^{0}$, D$_2^{*0}$, and 
background events
were found to be consistent with zero mean and unit variance for
both the $\Dzero$ decay modes taken separately as well as when fitting to the 
two modes combined.
The uncertainty is taken to be the error on the mean of the pull distributions.
We note that the two narrow peaks are resolvable
and we are able to correctly
fit the number of D$_{1}^{0}$ and D$_2^{*0}$ signal events
in the Monte Carlo simulations
for various input signal branching ratios.

\section{Results}
\label{s:results}

\begin{figure}[!p]
        \vspace{0.8cm}
        \begin{center}
            \resizebox{\linewidth}{!}{\includegraphics{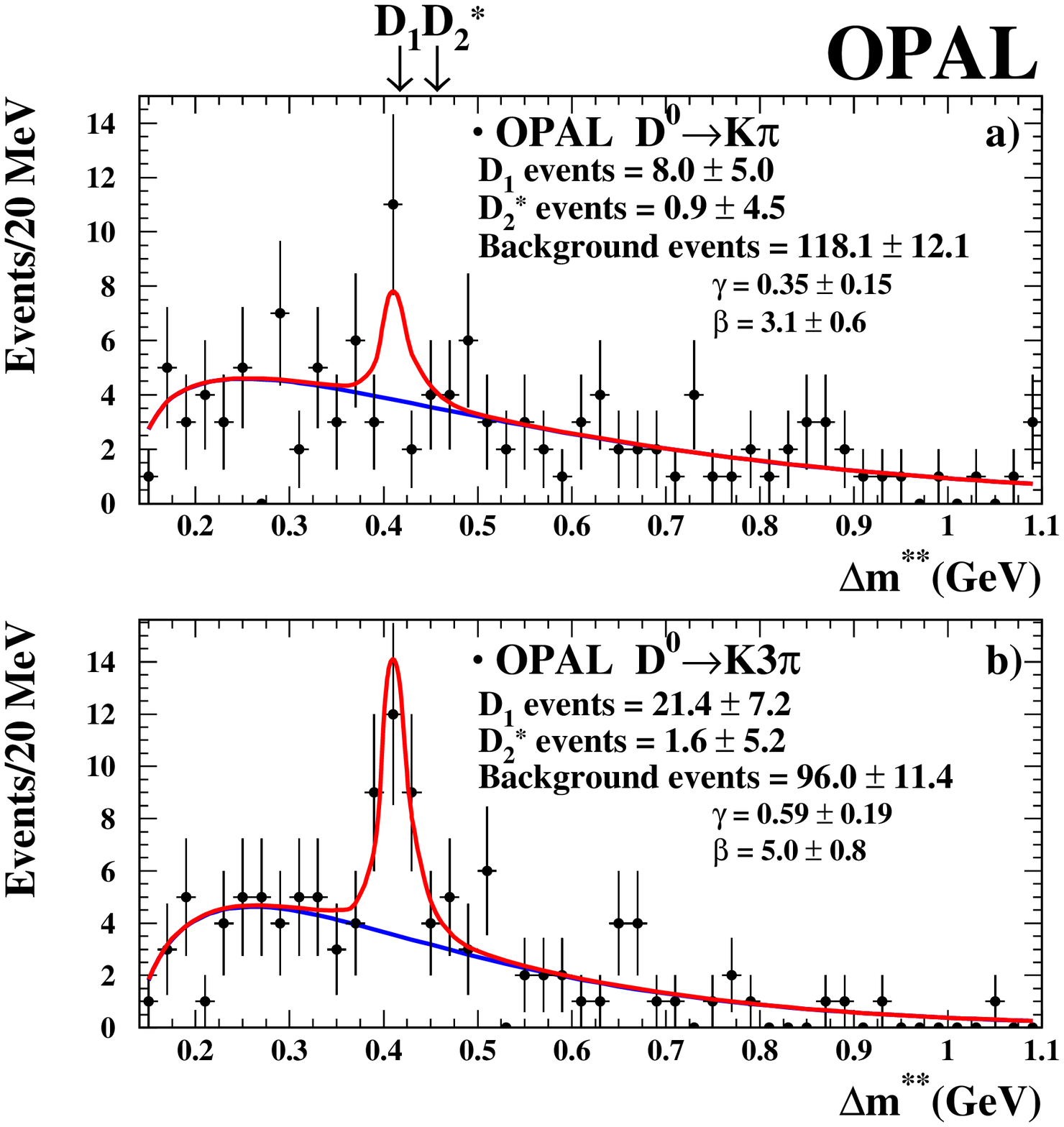} }
        \caption[resolution]{    
$\dssm$ distribution for 
a) the $\dkpi$ mode,  
b) the $\dktpi$ mode.
The superimposed lines show the overall fit 
and the background fitted shape. 
The expected positions of the D$_{1}^{0}$ and D$_2^{*0}$ are
indicated by the arrows.

        \label{f:kpik3pi} }
        \end{center}
    \end{figure}

The final $\dssm$ distributions for the $\dkpi$ and 
$\dktpi$ modes are shown in Fig.~\ref{f:kpik3pi}. 
The numerical results of the fit are also shown, including the $1\,\sigma$ uncertainties
on the fitted parameters.
A signal for the expected narrow D$_{1}^0$ peak is present
in each mode separately, with a weaker significance for the 
$\dkpi$ mode (8.0$\pm$5.0 fitted D$_{1}^{0}$ events) than for
$\dktpi$ (21.4$\pm$7.2 fitted D$_{1}^{0}$ events).
No evidence of a D$_{2}^{*0}$ signal is present. 

The efficiencies estimated for the D$_{1}^0$ state
using dedicated signal simulations
are found to be $7.7\pm0.4\%$ and $2.3\pm0.1\%$ for the $\dkpi$ and 
$\dktpi$ modes, respectively, where the errors are statistical only.
For the D$_{2}^{*0}$ state these are $9.2\pm0.8\%$ and $2.1\pm0.2\%$.
        The product branching ratio can then be obtained as:
\begin{eqnarray}
 \mathrm{BR}\left(\myb \rightarrow \bar{\mathrm{B}} \right) \times \BR\left(\bar{\mathrm{B}} \rightarrow \Dssz\, \ell^- \bar{\nu} X \right)\
\times \BR\left(\Dssz \rightarrow \Dstarp\pi^{-}\right) = \nonumber \\
\frac{N_{\mathrm{Fit}}(\Dssz)/\varepsilon_{\Dssz}}
{\left(N_{\mathrm{Z}} / \varepsilon_{\mathrm{Z}}\right)
\times R_\mathrm{b} \times 2 \times 2 \times \BR\left(\Dstarp \rightarrow \Dzero \pi^{+}\right) \times
\BR \left( \dktpicommakpi \right) } 
\end{eqnarray}
The two factors of 2 arise from the two $\myb$ hadrons in the event, and the
two flavors of tagged lepton.
The fraction, $R_\mathrm{b}$, of \Zbb events in \Z hadronic decays,  
and the branching ratios for $\Dstarp \rightarrow \Dzero \pi^{+}$ and 
$\dkpi$ 
are taken from~\cite{pdg02};
$N_{\mathrm{Z}}$ = \ntot\ and 
$\varepsilon_{\mathrm{Z}} = (98.1 \pm 0.5)\%$, as mentioned in 
section~\ref{s:presel}.

For decays into the D$_{1}^0$ state, we obtain for the $\dkpi$ mode:
\begin{eqnarray}
\mathrm{BR}\left(\myb \rightarrow \bar{\mathrm{B}} \right) \times \BR\left(\bar{\mathrm{B}} \rightarrow \mathrm{D}^{0}_1\, 
\ell^- \bar{\nu} X \right)\
\times \BR\left(\mathrm{D}^{0}_1 \rightarrow \Dstarp\pi^{-}\right)= 
 (1.17 \pm 0.73 \pm 0.27) \times 10^{-3} \nonumber
\end{eqnarray} and  for $\dktpi$:
\begin{eqnarray}
\mathrm{BR}\left(\myb \rightarrow \bar{\mathrm{B}} \right) \times \BR\left(\bar{\mathrm{B}} \rightarrow \mathrm{D}^{0}_1\, 
\ell^- \bar{\nu} X \right)\
\times \BR\left(\mathrm{D}^{0}_1 \rightarrow \Dstarp\pi^{-}\right)= 
(5.30 \pm 1.79 \pm 0.95) \times 10^{-3} \nonumber
\end{eqnarray} where the first error is statistical and the second 
systematic. The evaluation of the systematic uncertainty is discussed in section~\ref{s:syst}. The 
two product branching ratios measured for the $\dkpi$ and $\dktpi$ 
decay modes agree at the level of $1.9\,\sigma$.

\begin{figure}[!p]
        \vspace{0.8cm}
        \begin{center}
            \resizebox{\linewidth}{!}{\includegraphics{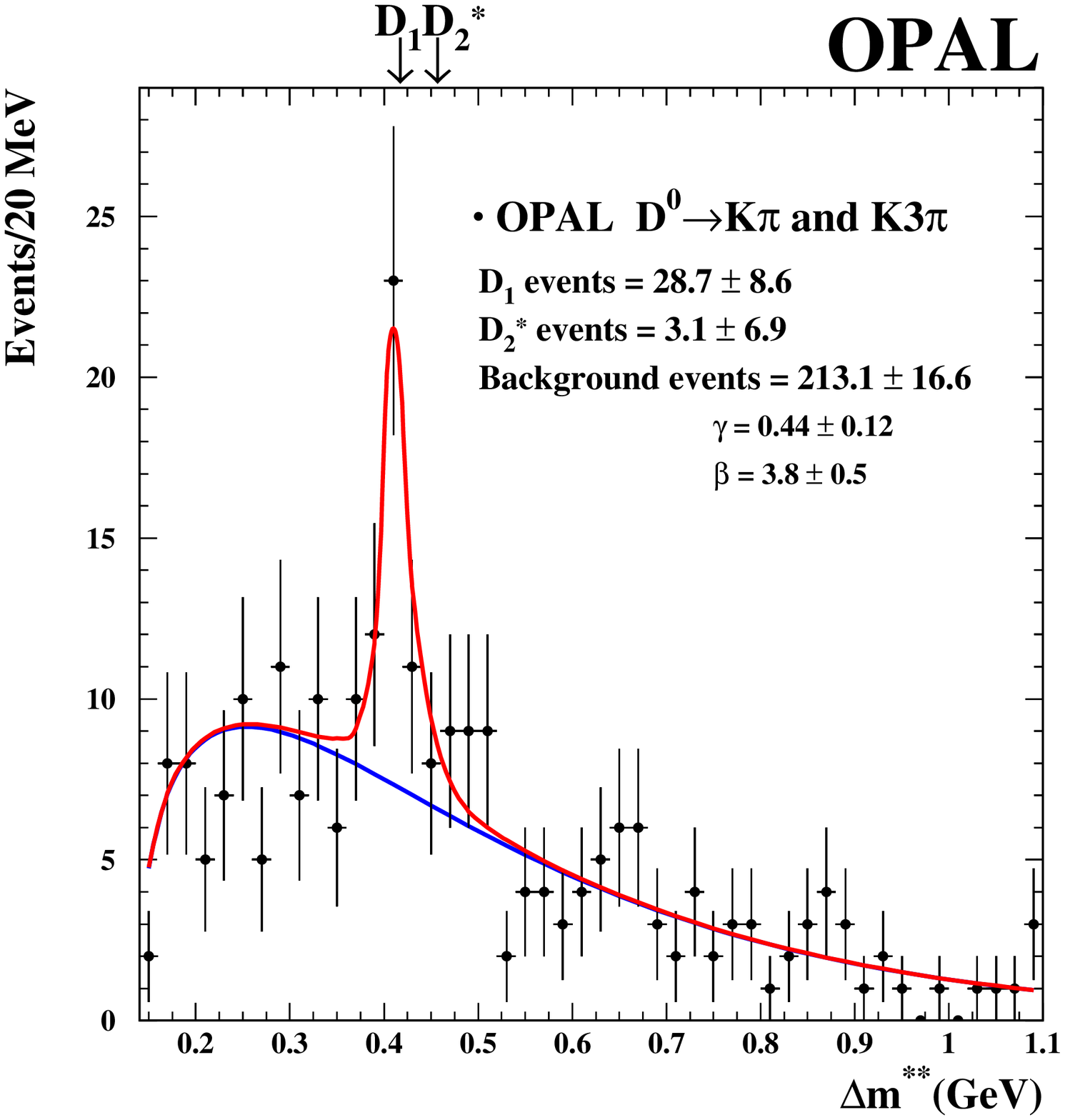} }
        \caption[resolution]{    
$\dssm$ distribution for $\dkpi$ and $\dktpi$ combined.
The superimposed lines show the overall fit 
and the background fitted shape. 
The expected positions of the D$_{1}^{0}$ and D$_2^{*0}$ are
indicated by the arrows.

        \label{f:combo} }
        \end{center}
    \end{figure}

The two modes were combined by merging the two data samples
and performing a single fit,
which has the advantage of reducing the systematic error on the background. 
The resulting distribution is shown in Fig~\ref{f:combo}. 
The combined efficiency,
multiplied by the respective $\Dzero$ decay branching ratios,
is 0.47\% $(7.7\%\times3.80\%+2.3\%\times7.46\%)$.
There are $28.7\pm8.6$ events in the D$_{1}^0$ peak, 
from which the product branching ratio follows:
\begin{eqnarray}
\mathrm{BR}\left(\myb \rightarrow \bar{\mathrm{B}} \right) \times \BR\left(\bar{\mathrm{B}} \rightarrow \mathrm{D}^{0}_1\, 
\ell^- \bar{\nu} X \right)\
\times \BR\left(\mathrm{D}^{0}_1 \rightarrow \Dstarp\pi^{-}\right)= 
 (2.64 \pm 0.79 \pm 0.39) \times 10^{-3}. \nonumber
\end{eqnarray}

The combined-sample fit yields
$3.1\pm6.9$ events for the D$_{2}^{*0}$.
The measured branching ratio is then
consistent with zero:
\begin{eqnarray}
\mathrm{BR}\left(\myb \rightarrow \bar{\mathrm{B}} \right) \times \BR\left(\bar{\mathrm{B}} \rightarrow \mathrm{D}^{*0}_2\, 
\ell^- \bar{\nu} X \right)\
\times \BR\left(\mathrm{D}^{*0}_2 \rightarrow \Dstarp\pi^{-}\right)= 
 (0.26 \pm 0.59 \pm 0.35) \times 10^{-3}. \nonumber
\end{eqnarray}
Using the method of Feldman and Cousins~\cite{ref:cousins,pdg02},
this can be converted into a 95\% confidence level upper limit: 
\begin{eqnarray}
\mathrm{BR}\left(\myb \rightarrow \bar{\mathrm{B}} \right) \times \BR\left(\bar{\mathrm{B}} \rightarrow \mathrm{D}^{*0}_2\, 
\ell^- \bar{\nu} X \right)\
\times \BR\left(\mathrm{D}^{*0}_2 \rightarrow \Dstarp\pi^{-}\right) <
 1.39 \times 10^{-3}. 
\nonumber
\end{eqnarray}

These results were checked in several ways.
Firstly, if instead of fixing the  D$_{1}^0$ mass
we include it as an extra parameter to be fitted
we obtain:
\begin{eqnarray}
\mathrm{BR}\left(\myb \rightarrow \bar{\mathrm{B}} \right) \times \BR\left(\bar{\mathrm{B}} \rightarrow \mathrm{D}^{0}_1\, 
\ell^- \bar{\nu} X \right)\
\times \BR\left(\mathrm{D}^{0}_1 \rightarrow \Dstarp\pi^{-}\right)= 
 (2.66 \pm 0.80 \pm 0.39) \times 10^{-3}, 
\nonumber
\end{eqnarray} 
in agreement with the previous result
of $(2.64 \pm 0.79 \pm 0.39) \times 10^{-3}$.
The fitted D$_{1}^0$ mass is ($2418.8\pm5.0$)~MeV, consistent with the world 
average of ($2422.2\pm1.9$)~MeV~\cite{pdg02}.
The cuts described in 
section~\ref{s:finalcuts} were varied by significant amounts 
(10\%--100\%),
and the resulting variations in the 
measured branching ratios were found to be compatible with the 
expected statistical variations.
To check the procedure for combining the results of the two $\Dzero$ decay modes, we
calculated a simple weighted average of the two results.
This average is consistent with the combined result.

\begin{figure}[!p]
        \vspace{0.8cm}
        \begin{center}
            \resizebox{\linewidth}{!}{\includegraphics{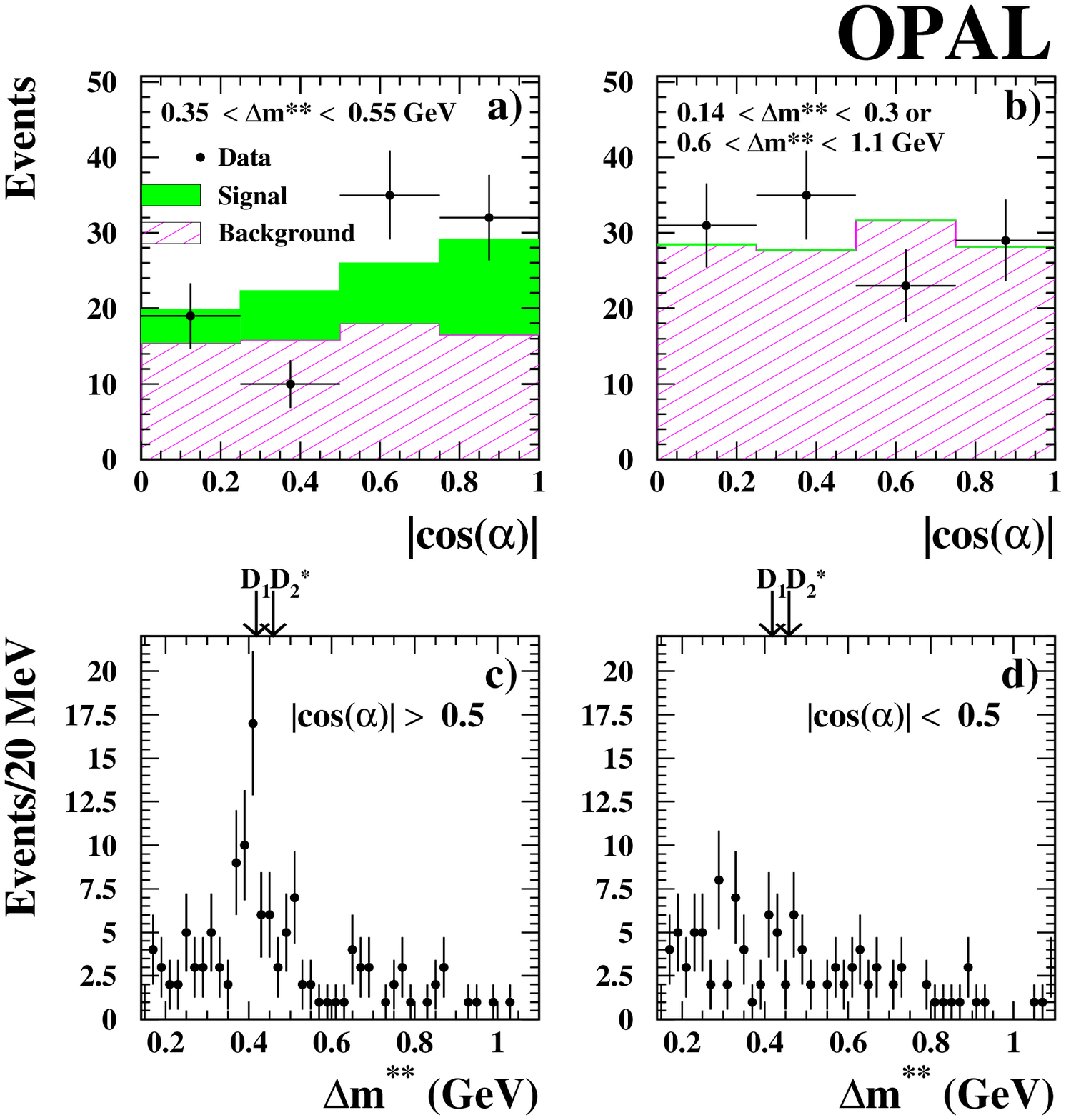} }
\caption[resolution]{
Distributions dependent on the decay angle $\alpha$:
a) $|\cos(\alpha)|$ for the combined $\dkpi$ and $\dktpi$ 
data and Monte Carlo
in the $0.35<\dssm<0.55$ GeV signal band;
b) complementary events to a) in the $0.14<\dssm<0.3$ and $0.6<\dssm<1.1$ GeV sidebands;
c) $\dssm$ data distributions for $\dkpi$ and $\dktpi$ combined data
in the $|\cos(\alpha)|>0.5$ region (expected to enhance the D$_{1}$ signal);
d) complementary events to c) for $|\cos(\alpha)|<0.5$ (expected to suppress the D$_{1}$ signal).
The expected positions of the D$_{1}$ and D$_2^{*}$ are indicated
by arrows.

        \label{f:alpha} }
        \end{center}
    \end{figure}

We also considered the angular decay distribution of the narrow states.
In the heavy-quark limit, the pure D-wave decay of the $J^{P}=2^{+}$ D$^{*0}_2$ state
should be distributed according to
 $\frac{3}{4}\sin^{2}\alpha$, whereas for the $J^{P}=1^{+}$ D$^{0}_1$ we expect
$\frac{1}{4}(1+3\cos^{2}\alpha)$~\cite{rosner}. This has been observed experimentally by
CLEO~\cite{cleodecay}.
Here $\alpha$ is the angle between the $\piss$ and the $\pislow$
from the D$^{*+}$ decay, evaluated in the rest frame of the D$^{*+}$.
For D$^{0}_1$ events we expect more events at larger values of $|\cos \alpha|$, 
and this is evident in Fig.~\ref{f:alpha}a which shows this angle 
for data and Monte Carlo in the signal mass region $0.35<\dssm<0.55$~GeV.
The shape of the data distribution is observed to agree with that expected from
simulated events, where the signal rate is fixed to the one measured here.
Fig.~\ref{f:alpha}b shows that background events, 
selected from right sign data in the sidebands 
$0.14<\dssm<0.3$ or $0.6<\dssm<1.1$~GeV,
exhibit a flat $\cos\alpha$ distribution.
An enhancement of the D$^{0}_1$ resonance peak is expected
for higher values of $|\cos \alpha|$, and
this is confirmed in Fig.~\ref{f:alpha}c
as the $\dssm$ D$^{0}_1$ peak is enhanced with a selection $|\cos \alpha|>0.5$. 
Note how the overall level of the background drops relative to
Fig.~\ref{f:combo}.
Conversely, for $|\cos \alpha|<0.5$ (Fig.~\ref{f:alpha}d), 
the D$^{0}_1$ peak greatly diminishes in significance. 

For a D$^{*0}_2$ enhancement selection, namely $|\cos \alpha|<0.2$, 
we find no distinct sharp peak at the expected D$^{*0}_2$ position. 
This provides additional evidence for the presence of the
D$_{1}^0$ state and absence of the D$^{*0}_2$.

\section{Systematic Uncertainties}
\label{s:syst}

The systematic uncertainties in the product branching ratios are shown in 
Table~\ref{tablesyst}. The dominant contributions arise from uncertainties in
the background function and the D$_{1}^0$ and D$^{*0}_2$ fit parameters.
The components of systematic uncertainty shown in the table are:

\begin{itemize}

\item Background function: 
The dominant uncertainty on the background arises
from the possibility that some background has a peak
in the signal region,
therefore biasing the fitted number of D$_{1}^0$ and D$^{*0}_2$ signal events. 
This uncertainty is estimated
from the average number of (expected-fitted) signal events, as obtained in the Monte Carlo
ensemble tests described in Section~\ref{s:fit}. 
There is also a much smaller background
uncertainty on the overall background shape.
This component of the uncertainty is estimated
by comparing the $\gamma$, $\beta$  parameters
obtained from Monte Carlo with those obtained from the wrong sign
$\piss$ data sample.

\item $\Dssz$ fit parameters:
We vary the D$_{1}^0$ and D$^{*0}_2$ masses and widths within the current uncertainties~\cite{pdg02},
and refit the $\dssm$ distribution. We also vary the Gaussian resolution by 
$\pm 1$ MeV and redo the fit. The resulting variations are added in quadrature.

\end{itemize}

All the remaining sources of systematic uncertainty affect only the signal efficiency, 
as the background shape and normalisation are obtained directly from the data.

\begin{itemize}

\item The limited Monte Carlo statistics in the signal samples give rise to a systematic uncertainty on the
estimation of efficiencies.

\item The lepton identification efficiency has an uncertainty of 3\% for muons and 4\% for 
electrons ~\cite{ref:rb}. 

\item The mean and sigma of the normalised Monte Carlo $\dedx$ distributions 
were varied by $\pm10\%$~\cite{ref:dedx}.

\item Tracking resolution:
The systematic uncertainty was assessed in Monte
Carlo by applying a global 10\% degradation to the resolution of all 
measured track parameters.
\item The lifetime of the $\bar{\mathrm{B}}$ mesons was varied within their measured uncertainty~\cite{pdg02}.

\item Theoretical uncertainty in the $\bjxn$ form factors:
Different theoretical models predict different form factors for the B semileptonic decay, 
and therefore different $\piss$ energy spectra. This source of systematic 
uncertainty is dominant at the 
$\Upsilon(4\mathrm{S})$ experiments~\cite{cleo}. Due to the high boost provided 
in \Zbb decays, this uncertainty is expected to be significantly
smaller in the analyses at LEP. This 
was confirmed by reweighting our JETSET~\cite{bib:jetset} Monte Carlo samples to the
forms factors described in~\cite{ligeti}. The range of parameters describing the
form factor calculation
was varied within the values constrained as in~\cite{richvcb}, and also 
the values needed to approximately reproduce the form factors predicted by~\cite{igsw2}.
The maximum variations in the signal efficiency were taken. 

\item B fragmentation:
The Peterson fragmentation model~\cite{bib:peter} is used in JETSET~\cite{bib:jetset}
to model the momentum distribution of the B hadrons. Our Monte Carlo samples were
reweighted event by event to reproduce the experimental uncertainty on the mean energy
of the B hadrons. The Peterson parameter $\varepsilon_\qb$ was varied within the range
obtained in~\cite{pr359}. 
The variations are in agreement with those obtained using two other fragmentation 
models~\cite{bib-kartvelishvili,bib-colspi,bib-bowler}, again with
parameter ranges as determined in~\cite{pr359}.

\item The relevant branching ratios and $R_\mathrm{b}$ were varied within the 
published uncertainties~\cite{pdg02}.

\item The hadronic Z preselection efficiency was varied by its uncertainty of $\pm 0.5\%$~\cite{bib:gambb}.

\end{itemize}

%
\section{Conclusions}

We have analysed semileptonic B decays into the narrow
D$_1^0$ and D$_2^{*0}$ states in events compatible with the decay chain
$\myb \rightarrow \bar{\mathrm{B}} \rightarrow \Dssz\,\ell^- \bar{\nu} X $, $\Dssz \rightarrow \Dstarp\pi^{-}$, $\dsd$, $\dktpiorkpi$.
The product branching ratio
for decays into the D$_1^0$ state is measured to be:
\begin{eqnarray}
\mathrm{BR}\left(\myb \rightarrow \bar{\mathrm{B}} \right) \times \BR\left(\bar{\mathrm{B}} \rightarrow \mathrm{D}^{0}_1\, 
\ell^- \bar{\nu} X \right)\
\times \BR\left(\mathrm{D}^{0}_1 \rightarrow \Dstarp\pi^{-}\right)= 
 (2.64 \pm 0.79 \pm 0.39) \times 10^{-3}, \nonumber
\end{eqnarray}
where the first error is statistical and the second systematic.
We find no evidence for decays into the 
$J^{P}=2^{+}$ state, and set a limit on the product branching ratio:
\begin{eqnarray}
\mathrm{BR}\left(\myb \rightarrow \bar{\mathrm{B}} \right) \times \BR\left(\bar{\mathrm{B}} \rightarrow \mathrm{D}^{*0}_2\, 
\ell^- \bar{\nu} X \right)\
\times \BR\left(\mathrm{D}^{*0}_2 \rightarrow \Dstarp\pi^{-}\right) <
 1.4 \times 10^{-3}, \nonumber
\end{eqnarray} at the 95\% confidence level.
These results update a previous OPAL analysis~\cite{oldopal}, and agree with 
similar measurements performed at 
LEP~\cite{aleph,delphi}, CLEO~\cite{cleo}, and ARGUS~\cite{argus}.


\bigskip\bigskip\bigskip
\begin{flushleft}
{\Large\bf Acknowledgements}
\end{flushleft}
 We particularly wish to thank the SL Division for the efficient operation
 of the LEP accelerator at all energies
 and for their close cooperation with
 our experimental group.  In addition to the support staff at our own
 institutions we are pleased to acknowledge  \\
 Support from the scholarship PRAXIS/BD/11460/97, Portugal, \\
 Department of Energy, USA, \\
 National Science Foundation, USA, \\
 Particle Physics and Astronomy Research Council, UK, \\
 Natural Sciences and Engineering Research Council, Canada, \\
 Israel Science Foundation, administered by the Israel
 Academy of Science and Humanities, \\
 Benoziyo Center for High Energy Physics,\\
 Japanese Ministry of Education, Culture, Sports, Science and
 Technology (MEXT) and a grant under the MEXT International
 Science Research Program,\\
 Japanese Society for the Promotion of Science (JSPS),\\
 German Israeli Bi-national Science Foundation (GIF), \\
 Bundesministerium f\"ur Bildung und Forschung, Germany, \\
 National Research Council of Canada, \\
 Hungarian Foundation for Scientific Research, OTKA T-029328, 
 and T-038240,\\
 The NWO/NATO Fund for Scientific Reasearch, the Netherlands.\\


\newpage
%
\begin{table}[htbp]
\begin{center}


\begin{tabular}{|l l l l l l|}\hline

 & & & Mass & Width & Decay Modes \\

 & $J^P$ & $J_{q}$  & (MeV) & (MeV) & \\ \hline

 D$_{1}^{'0}$ & 1$^{+}$ & 1/2 & $\sim$2470  & $\gsim$250  & D$^*\pi$ \\ 
 D$_{0}^{*0}$ & 0$^{+}$ & 1/2 & $\sim$2400  & $\gsim$170   & D$\pi$ \\ 
 D$_{1}^{0}(2420)$ & 1$^{+}$ & 3/2 & 2422.2 $\pm$ 1.8 & 18.9$^{\mathrm{+4.6}}_{\mathrm{-3.5}}$  & D$^*\pi$ \\ 
 D$_2^{*0}(2460)$ & 2$^{+}$ & 3/2 & 2458.9 $\pm$ 2.0 & 23 $\pm$ 5  & D$\pi$, D$^*\pi$ \\ \hline
\end{tabular}
\end{center}
  \caption[D$^{**}$ states]
   {Properties of neutral orbitally-excited charm mesons.
The quantum number $J_{q}$ is the total angular momentum of the
light constituents of the meson. Masses and widths of the narrow $J_{q}=$~3/2 states 
are experimentally determined~\cite{pdg02}. For the light $J_{q}=$~1/2 states, 
we quote theoretical estimates of the masses and widths~\cite{fishytheory}.}
  \label{tabledss}
\end{table}
%
\begin{table}[htbp]
\begin{center}

\begin{tabular}{|l c c c|} \hline & & & \\ 
          &  \multicolumn{3}{c|}{Systematic Uncertainty $(\times 10^{-3})$}  \\ 
 Source & $K\pi$ mode  & $K\,3 \pi$ mode & Combined \\ \hline

Background function              & $\pm$ 0.23 & $\pm$ 0.54 & $\pm$ 0.17  \\ 
$\mathrm{D}_1, \mathrm{D}_2\,$ fit parameters      & $^{+0.09}_{-0.06}$ & $^{+0.54}_{-0.42}$ &  $^{+0.22}_{-0.17}$ \\ 
Signal simulation statistics     & $\pm$ 0.06 & $\pm$ 0.20 & $\pm$ 0.12  \\ 
Lepton ID                        & $\pm$ 0.04 & $\pm$ 0.18 & $\pm$ 0.09  \\ 
\dedx                            & $\pm$ 0.05 & $\pm$ 0.26 & $\pm$ 0.11  \\ 
Tracking resolution              & $\pm$ 0.04 & $\pm$ 0.27 & $\pm$ 0.10  \\ 
B lifetime                       & $\pm$ 0.01 & $\pm$ 0.05 & $\pm$ 0.03  \\ 
Theoretical uncertainty          & $\pm$ 0.03 & $\pm$ 0.16 & $\pm$ 0.08  \\ 
B fragmentation                  & $\pm$ 0.04 & $\pm$ 0.26 & $\pm$ 0.10  \\ 
$R_\mathrm{b}$                            & $\pm$ 0.004 & $\pm$ 0.02 & $\pm$ 0.009  \\ 
$\mathrm{BR}(\Dstarp \rightarrow \Dzero \pi)$    & $\pm$ 0.008 & $\pm$ 0.04 & $\pm$ 0.02  \\ 
$\mathrm{BR}(\Dzero \rightarrow \mathrm{K} \pi)$      & $\pm$ 0.03  &            & $\pm$ 0.02  \\ 
$\mathrm{BR}(\Dzero \rightarrow \mathrm{K}\, 3 \pi)$  &             & $\pm$ 0.22 & $\pm$ 0.13  \\ 
Hadronic Z preselection              & $\pm$ 0.006 & $\pm$ 0.03 & $\pm$ 0.01  \\
 & & & \\
Total                            & $^{+0.27}_{-0.26}$ & $^{+0.98}_{-0.92}$ & $^{+0.40}_{-0.38}$  \\ 

\hline 

\end{tabular}

\end{center}
  \caption[Systematic errors]
   { Systematic uncertainties on the product branching ratios. The combined
column is for the combination resulting 
from merging the two data samples and performing a single fit.}
  \label{tablesyst}
\end{table}

\newpage


\newpage


\newpage


\newpage


\newpage



\newpage


\end{document}